# Enhanced mobility of high-frequency discrete breathers in a monatomic chain with odd anharmonicity.

A. Shelkan[1], M. Klopov[2], V. Hizhnyakov[1]

[1]*Institute of Physics, University of Tartu, Ravila 14c, 50411 Tartu, Estonia*
[2]*Department of Physics, Tallinn University of Technology, Ehitajate 5, 19086 Tallinn, Estonia*
shell@ut.ee

**Abstract.**
The mobility of high-frequency discrete breathers in monatomic chains with nonlinear interatomic potentials of the nearest neighbours is considered. It was found that the odd (cubic and fifth) anharmonicity strongly affects the mobility of breathers, sharply increasing the distance that it propagates without being trapped. It was also found that the correctly chosen fifth anharmonicity leads to an inversion of stability between the bond-centred and site-centers breathers and to the low-radiative propagation of discrete breathers along the chain.

1. **Introduction.**

Previous analytical studies of nonlinear dynamics and molecular dynamics (MD) simulations have shown that local vibrational excitations, both stationary and moving, can exist in ideal anharmonic atomic lattices. These excitations are called intrinsic localized modes or discrete breathers (DBs) [1-3]. Large-size low-frequency DBs in atomic chains are highly mobile. The reason for this mobility is that for large-size breathers, the lattice discreteness can be neglected and DBs can be considered in the limit of a continuous medium. In this limit, the equations of motion can be rewritten in a moving reference frame and they will still have self-localized solutions describing large-size excitations. For small-size high-frequency DBs the atomic lattice discreteness significantly breaks the continuous symmetry and leads to the appearance of new phenomena, one of which is the trapping of moving DBs by the discreteness. This effect, first discovered for kinks [4], is caused by the appearance of Peierls-Nabarro (PN) barrier - the minimum energy barrier that needs to be overcome to transfer the excitation over a lattice period. Unlikely to kinks, breathers also have internal oscillation modes which may increase or decrease their energy. PN barrier for DBs has not been found explicitly, but illustrated via computer simulations, showing the trapping of DBs when their size is small enough [5.6].

In this communication we investigate the mobility of DBs in monatomic chains with nearest neighbor intersite forces, taking into account both, the even and the odd anharmonicities. Unlike the even one, odd anharmonicity changes the equilibrium position of atoms, which contribute to the DB, increases the length of the bonds (DC-shifts) and softens the elastic springs. One could think that the appearance of DC components should impede the movement of the DBs, as is usual for self-trapped quasiparticles. However, it turned out that adding even a small third or fifth anharmonicity leads to a large increase in the mobility of the DB of a given size and frequency.

2. **Molecular dynamics simulations and results.**

In our study we performed calculations of moving DBs in the Fermi-Pasta-Ulam (FPU) chains with the following potential energy of the nearest neighbors:

$$V = \frac{1}{2}k_2 x_n^2 + \frac{1}{3}k_3 x_n^3 + \frac{1}{4}k_4 x_n^4 + \frac{1}{5}k_5 x_n^5 \qquad (1)$$

Here $x_n = u_n - u_{n-1}$ is the difference of the displacements of the particles $n$ and $n-1$ from their equilibrium sites (i.e. $x_n$ is the distortion of the length of the interatomic bond number $n$ from the unperturbed chain bond length value). We are using the displacement units, for which $k_2 = k_4 = 1$ and considering the dependence of the movability of small size DBs with the frequencies $\omega \approx 1.5\omega_m$, $\omega \approx 2.0\omega_m$ and $\omega \approx 2.5\omega_m$ on $k_3$ and $k_5$. Here $\omega_m = 2$ is the top phonon frequency in the chain.

In our calculations, we used the Verlet and the fourth-order Runge-Kutta methods with time steps less than $0.0002/\omega_m = 0.0001$. The MD calculations are performed in a chain with 4000 atoms and fixed ends. The DB starts moving from the center of the chain. When it moves over 10 sites, we put the DB and its surroundings (displacements and speeds of 2000 atoms) back into the center of the chain. At the same, time we smoothly remove the velocities of the first and last 500 atoms of the central area of atoms with numbers from 1000 and 3000. In these regions we also smoothly reduce the displacements to the mean DC-shift value of the chain with odd anharmonicity. This allows us to eliminate the effect on the DB of the phonons, emitted during the motion and then reflected by the ends of the chain. If the DB moves to less than 10,000 sites, we also perform calculations on long (more than 100,000 atoms) chains without using the above procedure. In these long chains, the emitted phonons do not affect the DB. Both calculations give similar results.

The moving DB with the internal frequency $\omega \approx 1.5\omega_m = 3$ can be generated by initial even-parity displacement pattern (...0.0, 0.3, -0.86, 0.86, -0.3, 0.0...) and the initial velocity pattern orthogonal to it (...0.0, -v, v, v, -v, 0.0...). The first pattern is taken so that it coincides with the coordinates of a stationary (immobile) DB of a given frequency; the second pattern is similar to the velocity pattern of the linear local mode (LLM) induced by this stationary DB [7]. If the initial velocity parameter v is infinitesimal, then one gets a stationary DB with LLM. The velocity parameter v ☐ 0.1 gives a DB, which moves between the two sites back and forth. And if v is 0.2 or more, the DB moves along the lattice from the one site to another, its displacement pattern alternating approximately between even and odd DB patterns. When the moving DB is situated between two sites its displacement and velocity patterns approximately coincide with the initial patterns. The kinetic energy power spectrum of the moving DB is similar to the DB+LLM spectrum (see Fig. 1).

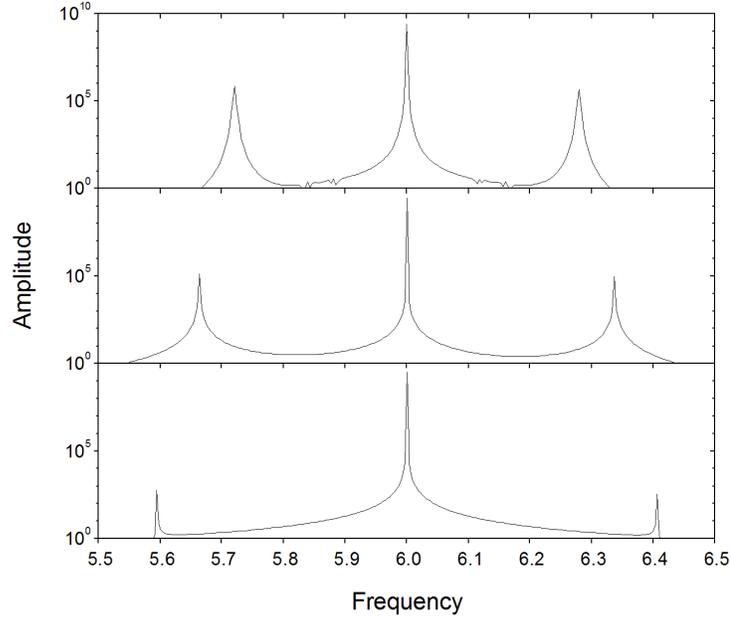

Fig.1. The power spectra of the kinetic energy of DBs with the frequency $\omega \approx 3$: stationary DB+LLM (down), DB oscillating between two cites (middle) and moving DB (up). Oscillating between two sites and moving DB spectra have the satellites similar to LLM (note, that kinetic energy oscillates with twice higher frequency than velocity).

We are considering the initial velocity patterns with different velocity parameters v = 0.2, 0.3, 0.4 and 0.5. If the DBs move in the chain without odd anharmonicity ($k_3 = 0$ and $k_5 = 0$) they stop between 4000 and 10000 atoms (see analogous also in [8,9]). But already in the chain with very small odd anharmonicity ($k_3 = 0.01$ or $k_5 = 0.01$) the mobility of DBs increases drastically and they travel across more than million sites without trapping.

To get a moving DB with the frequency $\omega \approx 4$ in the lattice with only quartic anharmonicity we start with the initial displacements (...0.0, 0.33, -1.33, 1.33, -0.33, 0.0...) and initial velocities (...0.0, -v, v, v, -v, 0.0...). The mobility of this DB is less, than that of the DB with $\omega \approx 3$. In the chain without an odd anharmonicity we considered 80 values of v between v = 0 and v = 0.8. In the down panel of Fig. 2 we present the number of the sites that the DB travels across during $t \approx 6000/\omega_m = 3000$. We found that all moving DBs were trapped earlier than site 200. E.g. if v = 0.5 then the DB with $\omega \approx 4$ moves over 38 sites and then it is trapped (see Fig. 3).

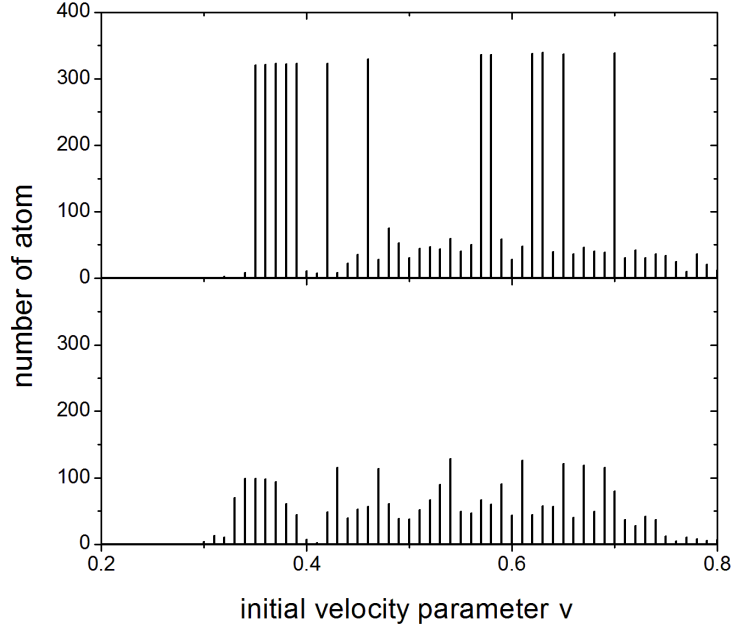

Fig. 2. The dependence of the traveling distance of DB with initial frequency $\omega \approx 4$ during $t \approx 6000/\omega_m = 3000$ on the initial velocity parameter $v$ in FPU chain with $k_3 = 0$, $k_5 = 0$ (down) and $k_3 = 0.15$, $k_5 = 0$ (up).

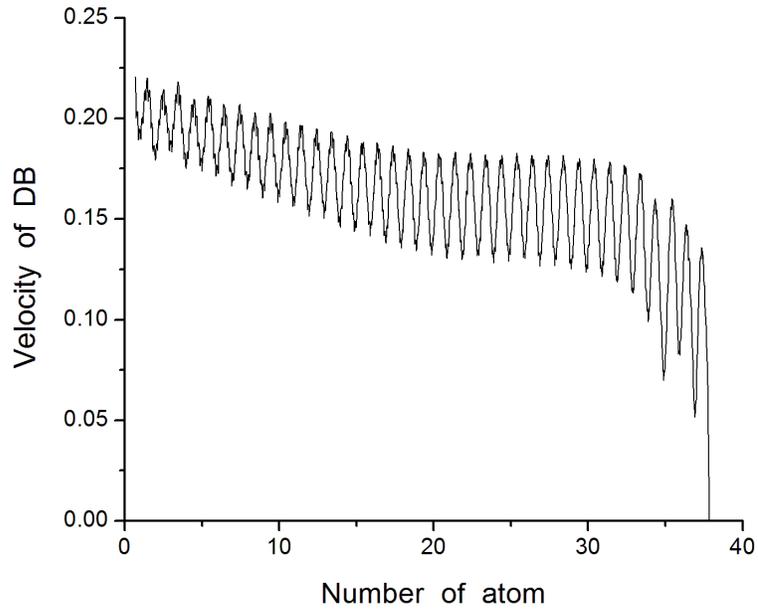

Fig. 3. Dependence of the mean velocity of moving DB with the initial frequency $\omega \approx 2\omega_m = 4$ on the traveling distance in FPU chain with $k_2 = k_4 = 1$ and $k_3 = k_5 = 0$.

In this figure one can also see that the mean (averaged over the DB oscillation period) velocity of the DB is larger when the DB is situated between the sites and smaller when it is situated on the site. This may be explained by the existence of a PN barrier. However, when we include an odd anharmonic term into the interaction potential ($k_3 = 0.15$ and larger or $k_5 = 0.1$ and larger), some of the DBs with $\omega \approx 4$ propagate over much longer distances without trapping. In the upper panel of Fig. 2 we can see, that in the chain with $k_3 = 0.15$ for some initial v the DBs with displacement pattern (...0.0, 0.3, -1.35, 1.35, -0.3, 0.0...) can travel over more than 300 sites during $t \approx 3000$. Our calculations show that, in fact these DBs can travel over more than 100,000 atomic sites (see Fig. 4, where the initial velocity pattern is v=0.35)

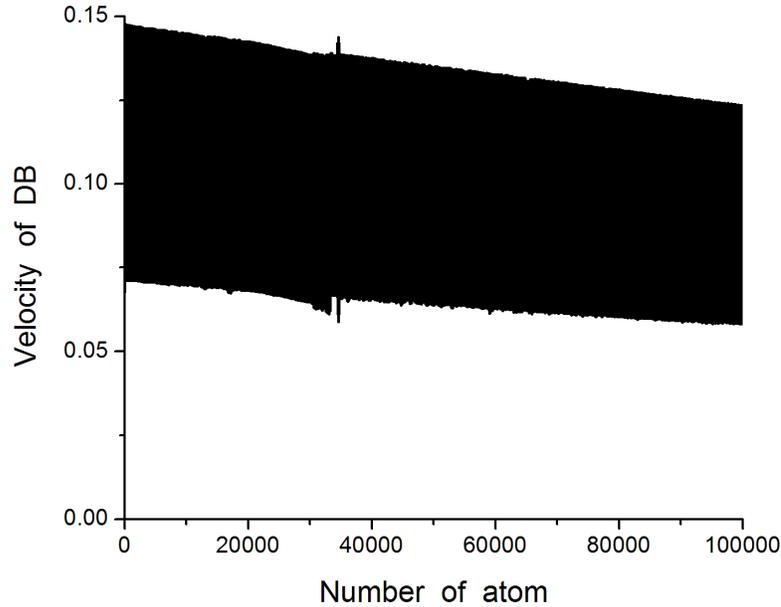

Fig. 4. The same as in Fig. 3 but in the case of $k_3 = 0.15$. The black area is caused by the oscillation of the mean DB velocity.

Third anharmonicity leads to the increasing of the DB mobility, but the higher the frequency of DB is the larger $k_3$ is needed to make the increase in mobility appreciable. The DBs with the initial frequency $\omega \approx 5$ travel over more than 100,000 sites if $k_3 = 0.3$ or larger. The distance covered by the DBs with the same initial frequency before trapping in the chain with zero odd anharmonicities is 1000 times smaller.

An increase in the third anharmonicity leads to an increase in the radiation of phonons by the moving DB and reducing of its internal energy and frequency. The dependence of the frequencies of DBs on propagating distance for different $k_3$ is presented in Fig. 5 (a), (b) and (c). The frequency change (diminishing) is a result of the DB internal energy loss. If $k_3$ is larger than 0.1, it is visible, but DBs remain high-frequency and strongly localized over long time and distances.

An increase in the fifth anharmonicity can lead not only to an increase, but also to a decrease in radiation of phonons by the moving DB due to the inversion of stability. An analogous effect was reported for a Klein-Gordon model with double-well on-site potential [1] and for some other systems [9,10]. In FPU chains bond-centered DBs are stable, and site-centered DBs unstable [11]. According to our calculations, instability in a FPU chain occurs for the DBs with frequencies 3,4,5 centered between the sites, if $k_3 = 0$ and $k_5$ is 0.3, 0.18 and 0.14, respectively. Thereat, the lower the frequency of DB is the larger is $k_5$ when a stability inversion takes place. In the vicinity of the inversion point the difference between site-centered and bond-centered stationary DB energies is small, the DB propagates essentially radiationless and its energy and frequency remain unchanged (see Fig 5 (d) and (e) for the DBs with frequencies 4 and 5).

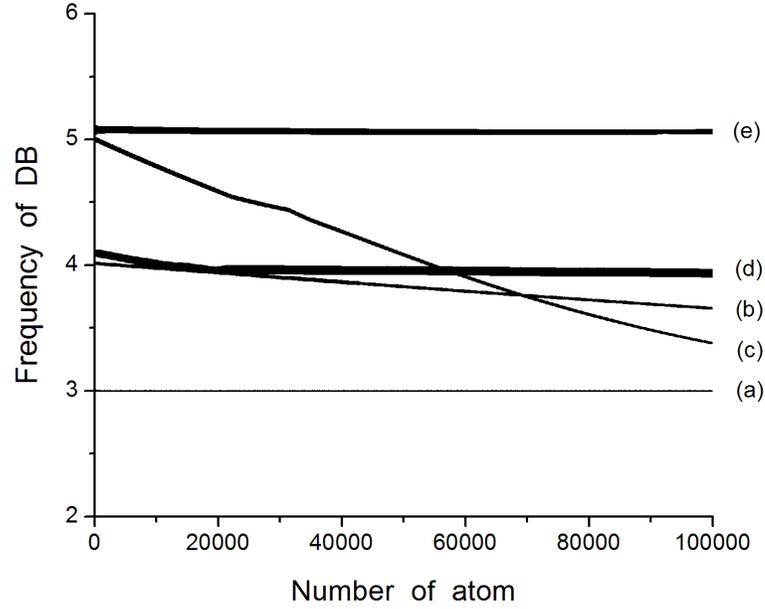

Fig. 5. The dependence of DB frequency on propagating distance: (a) $k_3 = 0.01$ $k_5 = 0$; (b) $k_3 = 0.15$, $k_5 = 0$; (c) $k_3 = 0.3$ $k_5 = 0$, (d) $k_3 = 0$, $k_5 = 0.18$; (e) $k_3 = 0$, $k_5 = 0.14$.

In the chain with $k_3 = 0$ and $k_5 = 0.18$ a low-radiation moving DB with the frequency $\omega \approx 4$ can be generated by exciting an initially immobile bond-centered DB with the frequency slightly higher than 4. After 300 oscillations the immobile DB with the frequency $\omega \approx 4.1$ and the initial displacement pattern (…-0.3, 0.05, -1.7, 1.7, -0.05, 0,3…) centered between the sites, loses its symmetry due to numerical inaccuracies and starts to move along the chain. The oscillations of the mean velocity of the moving DB are small (see Fig. 6).

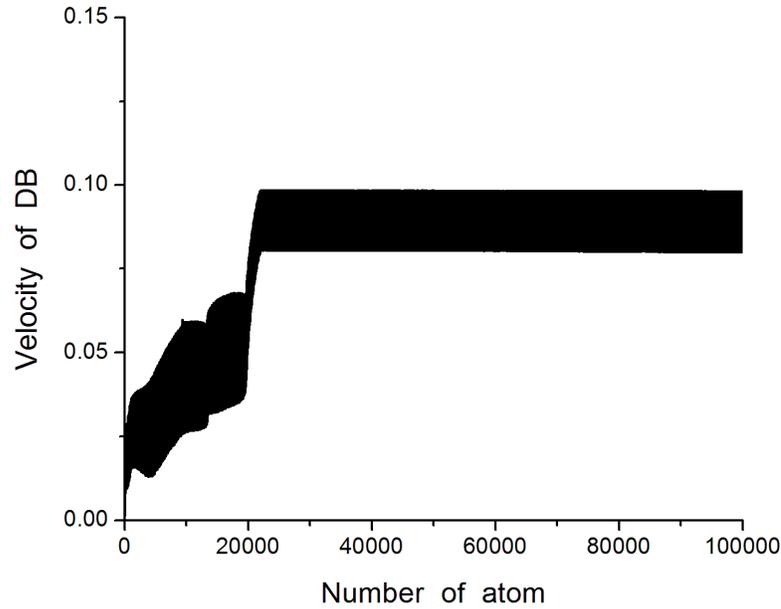

Fig.6. The dependence of the mean velocity of initially immobile unstable bond-centered DB with the frequency $\omega \approx 4$ in FPU chain with $k_3 = 0$, $k_5 = 0.18$ on the DB propagation distance.

The radiationless DBs can also propagate in a periodic 4000 atom chain (ring) with the fifth anharmonicity. Moreover, the DB can cover large distances also in a periodic chain with the third anharmonicity, if we exclude the influence of the phonons emitted during the motion. This can be done by reducing the displacements and velocities of 1000 atoms on the opposite to DB side of the ring on 0.0001% at each stage of calculation.

### 3. Conclusion

To sum up, in this communication we considered the mobility of high-frequency strongly-localized discrete breathers in monatomic chains with nearest-neighbor nonlinear interatomic potentials. We presented here the results of MD calculations of these breathers, showing their trapping by the Peierls-Nabarro barrier. We have found that odd (cubic and fifth) anharmonicity strongly affects the mobility of breathers drastically increasing the distance it propagates without trapping. We have also found that a properly chosen fifth anharmonicity leads to the stability inversion between bond-centered and site-centered stationary DBs and to essentially radiationless propagation of a DB along the chain. Weak satellites were observed in kinetic energy spectra of all moving high-frequency discrete breathers.

**References**

1. A.J. Sievers, S. Takeno, Phys. Rev. Lett. **61**, 970 (1988).
2. D. K. Campbell and M. Peyrard, in Chaos, edited by D. K. Campbell (AIP, New-York, 1990), p. 305.
3. S. Flach and C. R. Willis, Phys. Repts. **295**, 181 (1998).
4. M. Peyrard and M. D. Kruskal, Physica D **14**, 88 (1984).

5. T. Dauxois, M. Peyrard, and C. R. Willis, Phys. Rev. E **48**, 4768 (1993).
6. O. Bang and M. Peyrard, Physica D **81**, 9 (1995).
7. V. Hizhnyakov, A. Shelkan, M. Klopov, S. A. Kiselev, and A. J. Sievers, Phys. Rev. B **73**, 224302 (2006).
8. S. Aubry, Physica D **216**, 1 (2006).
9. J. L. Marín, S. Aubry, and L. M. Floría, Physica D **113**, 283 (1998).
10. M. Öster, M. Johansson, and A. Eriksson, Phys. Rev. E **67**, 056606 (1993).
11. K. W. Sandusky, J. B. Page, and K. E. Schmidt, Phys. Rev. B **46**, 6161 (1992).